\documentclass[12pt,a4paper]{iopart}
\usepackage{iopams}  
\usepackage{bm}
\begin{document}

\newcommand{\be}{\begin{equation}}
\newcommand{\ee}{\end{equation}}
\newcommand{\ba}{\begin{eqnarray}}
\newcommand{\ea}{\end{eqnarray}}
\def\D{{\mathcal D}{}}
\def\Gammabol{{\stackrel{\circ}{\Gamma}}{}}
\def\Abol{{\stackrel{~\circ}{A}}{}}
\def\Bbol{{\stackrel{~\circ}{B}}{}}
\def\Rbol{{\stackrel{\circ}{R}}{}}
\def\Obol{{\stackrel{\circ}{\Omega}}{}}
\def\Gammabol{{\stackrel{\circ}{\Gamma}}{}}
\def\Tbol{{\stackrel{\circ}{T}}{}}
\def\Dbol{{\stackrel{\circ}{\mathcal D}}{}}
\def\nabol{{\stackrel{\circ}{\nabla}}{}}
\def\Lbol{{\stackrel{\circ}{\mathcal L}}{}}
\def\Sbol{{\stackrel{\circ}{\mathcal S}}{}}
\def\tbol{{\stackrel{\circ}{t}}{}}
\def\Gammaw{{\stackrel{\bullet}{\Gamma}}{}}
\def\Omegaw{{\stackrel{\bullet}{\Omega}}{}}
\def\Aw{{\stackrel{~\bullet}{A}}{}}
\def\Vw{{\stackrel{\bullet}{V}}{}}
\def\Rw{{\stackrel{\bullet}{R}}{}}
\def\Qw{{\stackrel{\bullet}{Q}}{}}
\def\jw{{\stackrel{\bullet}{\jmath}}{}} 
\def\tw{{\stackrel{\bullet}{t}}{}}
\def\L{{\mathcal L}{}}
\def\Lw{{\stackrel{\bullet}{\mathcal L}}{}}
\def\Tw{{\stackrel{\bullet}{T}}{}}
\def\Kw{{\stackrel{\bullet}{K}}{}}
\def\nablaw{{\stackrel{\bullet}{\nabla}}{}}
\def\Dw{{\stackrel{\bullet}{\mathcal D}}{}}
\def\dw{{\stackrel{\bullet}{D}}{}}
\def\Sw{{\stackrel{\bullet}{\mathcal S}}{}}
\def\onehalf{{\textstyle{\frac{1}{2}}}}
\def\ihalf{{\textstyle{\frac{i}{2}}}}

\title[]{Hodge dual for soldered bundles}

\author{Tiago Gribl Lucas\footnote{E-mail: gribl@ift.unesp.br} and J. G. Pereira\footnote{E-mail: jpereira@ift.unesp.br}}

\address{Instituto de F\'{\i}sica Te\'orica, 
Universidade Estadual Paulista, Rua Pamplona 145, 01405-900 S\~ao 
Paulo, Brazil}

\begin{abstract}
In order to account for all possible contractions allowed by the presence of the solder form, a generalized Hodge dual is defined for the case of soldered bundles. Although for curvature the generalized dual coincides with the usual one, for torsion it gives a completely new dual definition. Starting from the standard form of a gauge lagrangian for the translation group, the generalized Hodge dual yields precisely the lagrangian of the teleparallel equivalent of general relativity, and consequently also the Einstein-Hilbert lagrangian of general relativity.
\end{abstract}




\section{Introduction}
\label{notation}

The geometrical setting of any gravitational theory is the tangent bundle, a natural
construction always present in spacetime. According to this structure, at each point of spacetime --- the base space of the bundle --- there is a tangent space attached to it --- the fiber of the
bundle --- on which the gauge group acts.\footnote{We use the Greek alphabet $(\mu, \nu, \rho, \dots = 0,1,2,3)$ to denote indices related to spacetime, and the Latin alphabet $(a,b,c, \dots = 0,1,2,3)$ to denote algebraic indices related to the tangent space, assumed to be a Minkowski spacetime with the metric $\eta_{ab}=\mathrm{diag}(+1,-1,-1,-1)$.} Differently from {\em internal} bundles of the Yang-Mills type gauge theories, spacetime-rooted bundles, as for example the tangent bundle, have a quite peculiar property: the presence of the solder form, whose components are the tetrad field \cite{koba}. For this reason, they are called soldered bundles. An immediate consequence of this property is that the connections living in these bundles will present, in addition to curvature, also torsion. This is the case, for example, of the Levi-Civita connection of general relativity, which has vanishing torsion.\footnote{We remark that the presence of a vanishing torsion is completely different from absence of torsion, which is the case of the non-soldered bundles of internal (or Yang-Mills) gauge theories.}

We denote the spacetime coordinates by $x^\mu$, whereas the tangent space co\-or\-di\-nates will be denoted by $x^a$. Since $x^a$ are functions of $x^\mu$, we can define coordinate basis for vector fields and their duals in the form
\be
\partial_a =
\left(\partial^\mu x_a \right) \partial_\mu \quad \mathrm{and} \quad 
\partial^a = \left(\partial_\mu x^a \right) \rmd x^\mu.
\label{1}
\ee
In these expressions, $\partial_{\mu} x^a$ is a trivial --- that is, holonomic --- tetrad, with $\partial^\mu x_a$ its inverse. A nontrivial tetrad field, on the other hand, defines naturally a non-coordinate basis for vector fields and
their duals,
\be \label{ninem}
h_a = h_a{}^\mu \partial_\mu \quad \mathrm{and} \quad h^a = h^a{}_\mu \rmd x^\mu.
\ee
These basis are non-holonomic,
\be \label{tenm}
[h_c , h_d]=f^a{}_{cd} \, h_a,
\ee
with
\be
f^a{}_{cd} = h_c{}^\mu \, h_d{}^\nu (\partial_\nu
h^a{}_\mu - \partial_\mu h^a{}_\nu)
\ee
the coefficient of anholonomy. A fundamental property of soldered bundles is that the spacetime (external) and the tangent--space (internal) metrics are related by
\be
g_{\mu\nu} = \eta_{ab} \, h^a{}_\mu \, h^b{}_\nu.
\label{gmn0}
\ee

A spin connection $A_\mu$ is a connection assuming values in the Lie
algebra of the Lorentz group,
\be
A_\mu = \onehalf \, A^{ab}{}_\mu \, S_{ab},
\ee
with $S_{ab}$ a given representation of the Lorentz generators. The corresponding covariant derivative is the Fock--Ivanenko operator \cite{fi,dirac}
\be
\D_\mu = \partial_\mu - \ihalf \, A^{ab}{}_\mu \, S_{ab}.
\ee
Acting on a Lorentz vector field $\phi^a$, for example, $S_{ab}$ is the matrix \cite{ramond}
\[
(S_{ab})^c{}_d = i \left(\delta_a{}^c \, \eta_{bd} - \delta_b{}^c \, \eta_{ad} \right),
\]
and consequently
\be
\D_\mu \phi^a = \partial_\mu \phi^a + A^{a}{}_{b \mu} \, \phi^b.
\ee
The spacetime linear connection $\Gamma^{\rho}{}_{\nu \mu}$ corresponding to $A^{a}{}_{b \mu}$ is
\be
\Gamma^{\rho}{}_{\nu \mu} = h_{a}{}^{\rho} \partial_{\mu} h^{a}{}_{\nu} +
h_{a}{}^{\rho} A^{a}{}_{b \mu} h^{b}{}_{\nu} \equiv h_{a}{}^{\rho} \D_{\mu} h^{a}{}_{\nu}.
\label{geco}
\ee
The inverse relation is
\be
A^{a}{}_{b \mu} =
h^{a}{}_{\nu} \partial_{\mu}  h_{b}{}^{\nu} +
h^{a}{}_{\nu} \Gamma^{\nu}{}_{\rho \mu} h_{b}{}^{\rho} \equiv h^{a}{}_{\nu} \nabla_{\mu} h_{b}{}^{\nu}.
\label{gsc}
\ee
Equations (\ref{geco}) and (\ref{gsc}) are different ways of expressing the property
that the total covariant derivative --- that is, with connection term for both indices --- of the tetrad vanishes
identically:
\be
\partial_{\mu} h^{a}{}_{\nu} - \Gamma^{\rho}{}_{\nu \mu} h^{a}{}_{\rho} +
A^{a}{}_{b \mu} h^{b}{}_{\nu} = 0.
\label{todete}
\ee

From a formal point of view, curvature and torsion are properties of connections. This becomes evident if we observe that many connections, with different curvature and torsion, are allowed to exist in the very same metric spacetime \cite{livro}. Given a connection $A^{a}{}_{b \mu}$, its curvature and the torsion are defined respectively by
\be
R^a{}_{b \nu \mu} = \partial_{\nu} A^{a}{}_{b \mu} -
\partial_{\mu} A^{a}{}_{b \nu} + A^a{}_{e \nu} A^e{}_{b \mu}
- A^a{}_{e \mu} A^e{}_{b \nu}
\ee
and
\be
T^a{}_{\nu \mu} = \partial_{\nu} h^{a}{}_{\mu} -
\partial_{\mu} h^{a}{}_{\nu} + A^a{}_{e \nu} h^e{}_{\mu}
- A^a{}_{e \mu} h^e{}_{\nu}.
\ee
Using relation (\ref{gsc}), they can be expressed in a purely spacetime form:
\be
\label{sixbm}
R^\rho{}_{\lambda\nu\mu} = \partial_\nu \Gamma^\rho{}_{\lambda \mu} -
\partial_\mu \Gamma^\rho{}_{\lambda \nu} +
\Gamma^\rho{}_{\eta \nu} \Gamma^\eta{}_{\lambda \mu} -
\Gamma^\rho{}_{\eta \mu} \Gamma^\eta{}_{\lambda \nu}
\ee
and
\be \label{sixam}
T^\rho{}_{\nu \mu} =
\Gamma^\rho{}_{\mu\nu}-\Gamma^\rho{}_{\nu\mu}.
\ee
The connection coefficients can be decomposed according to\footnote{All magnitudes related with general relativity will be denoted with an over ``$\circ$''.}
\be
\Gamma^\rho{}_{\mu\nu} = {\stackrel{\circ}{\Gamma}}{}^{\rho}{}_{\mu \nu} +
K^\rho{}_{\mu\nu},
\label{prela0}
\ee
where
\be
{\stackrel{\circ}{\Gamma}}{}^{\sigma}{}_{\mu \nu} = {\textstyle
\frac{1}{2}} g^{\sigma \rho} \left( \partial_{\mu} g_{\rho \nu} +
\partial_{\nu} g_{\rho \mu} - \partial_{\rho} g_{\mu \nu} \right)
\label{lci}
\ee
is the Levi--Civita connection of general relativity, and
\be
K^\rho{}_{\mu\nu} = {\textstyle
\frac{1}{2}} \left(T_\nu{}^\rho{}_\mu+T_\mu{}^\rho{}_\nu-
T^\rho{}_{\mu\nu}\right)
\label{contor}
\ee
is the contortion tensor.
Using relation (\ref{geco}), the decomposition (\ref{prela0}) can be rewritten as
\be
A^c{}_{a\nu} = \Abol^c{}_{a\nu} + K^c{}_{a\nu},
\label{rela00}
\ee
where $\Abol^c{}_{a \nu}$ is the Ricci coefficient of rotation, the spin connection of
general relativity.

\section{Dual operation for soldered bundles}

\subsection{General notions}

Let $\Omega^p$ be the space of $p$-forms on an $n$-dimensional manifold $M$. Since the vector spaces $\Omega^p$ and $\Omega^{n-p}$ have the same dimension, they are isomorphic. The choice of an orientation and the presence of a metric on $TM$ then enables us to single out a unique isomorphism, the so called Hodge dual \cite{baez}. For a $p$-form $\alpha^p \in \Omega^p$,
\be
\alpha^p = \frac{1}{p!} \, \alpha_{\mu_1 \dots \mu_p} \; \omega^{\mu_1} \wedge \dots \wedge \omega^{\mu_p},
\ee
its Hodge dual is the $(n-p)$-form $\star \, \alpha^p \in \Omega^{n-p}$ defined by
\be
\star \, \alpha^p = \frac{h}{(n-p)! p!} \, \epsilon_{\mu_1 \mu_2 \dots \mu_n} \, \alpha^{\mu_1 \dots \mu_p} \; \omega^{\mu_{p+1}} \wedge \dots \wedge \omega^{\mu_n}.
\ee
where we have used the identification $h = \sqrt{-g}$, with $h = \det (h^a{}_{\mu})$ and $g = \det (g_{\mu \nu})$. 
The operator $\star$ satisfies the property
\be
\star \star \alpha^p = (-1)^{p(n-p)+(n-s)/2} \alpha^p,
\label{property}
\ee	
where $s$ is the signature of the spacetime metric. Its inverse is given by
\be
\star^{-1} = (-1)^{p(n-p)+(n-s)/2} \star.
\ee

\subsection{The case of non-soldered bundles}

For non-soldered bundles, the dual operator can be defined in a straightforward way to act on vector-valued $p$-forms. Let $\beta$ be a vector-valued $p$-form on the $n$-dimensional base space $M$, taking values on a vector space $F$. Its dual is the vector-valued $(n-p)$-form
\be
\star \, \beta^p = \frac{h}{(n-p)! p!} \, \epsilon_{\mu_1 \mu_2 \dots \mu_n} \, e_i \, \beta^{i \, \mu_1 \dots \mu_p} \, \omega^{\mu_{p+1}} \wedge \dots \wedge \omega^{\mu_n},
\ee
where the set $\{ e_i \}$ is a basis for the vector space $F$. In this case, the components $\beta^{i \, \mu_1 \dots \mu_p}$ have also an internal space index $i$, which is not related to the external indices $\mu_i$. The property (\ref{property}) is of course still valid. As an example, let us consider the Yang-Mills field strength $F^A{}_{\mu \nu}$ in a four-dimensional spacetime. As the algebraic indices ($A, B, \dots$) are not related  to the spacetime indices ($\mu, \nu, \dots$), the Hodge dual is defined by \cite{frankel}
\be
\star F^A{}_{\mu \nu} = \frac{h}{2} \, \epsilon_{\mu \nu \rho \sigma} \, F^{A \rho \sigma}.
\label{*}
\ee

\subsection{The case of soldered bundles}

The case of soldered bundles is quite different. Due to the presence of the solder form, internal and external indices can be transformed into each other, and this feature leads to the possibility of defining new dual operators, each one related to an inner product on $\Omega^p$. The main requirement of these new definitions it that, since (\ref{property}) is still valid for $p$-forms on soldered bundles, we want to make it true also for vector-valued $p$-forms. We consider next, in a four-dimensional spacetime, the specific cases of torsion and curvature.

\subsubsection{Torsion}

Differently from internal (non-soldered) gauge theories, whose dual is defined by equation~(\ref{*}), in soldered bundles algebraic and spacetime indices can be transformed into each other through the use of the tetrad field. This property opens up the possibility of new contractions in relation to the usual definition (\ref{*}). There are basically two different kind of terms that can be taken into account when defining a generalized dual torsion. They are given by
\be
\star T^\lambda{}_{\mu \nu} = h \, \epsilon_{\mu \nu \rho \sigma} \left[
a \left(\onehalf \, T^{\lambda \rho \sigma} + T^{\rho \lambda \sigma}\right)
+ b \, T^{\theta \rho}{}_\theta \, g^{\lambda \sigma} \right],
\label{galstar1}
\ee
with $a, b$ constant coefficients.\footnote{See Ap\-pendix~A for a proof that two coefficients suffice to define the generalized dual torsion.} The factor $1/2$ in the first term is necessary to remove equivalent terms of the summation. Now, in a four--dimensional spacetime with metric signature $s = 2$, the dual torsion must satisfy the relation
\be
{\star \star}T^\rho{}_{\mu \nu} = - T^\rho{}_{\mu \nu}.
\label{relation}
\ee
This condition yields the following algebraic system:
\ba
2 a^2 - a b = 1 \label{ae1} \\
2 a^2 + a b = 0. \label{ae2}
\ea
There are two solutions which differ by a global sign:
\be
a = 1/2 \qquad b = - 1
\label{solution1}
\ee
and
\be
a = - 1/2 \qquad b = 1
\label{solution2}
\ee
Since we are looking for a generalization of the usual expression (\ref{*}), we choose
the solution with $a > 0$. In this case, the generalized dual torsion reads
\be
\star T^\rho{}_{\mu \nu} = h \, \epsilon_{\mu \nu \alpha \beta} \left(\textstyle{\frac{1}{4}} \, T^{\rho \alpha \beta} + \frac{1}{2} \, T^{\alpha  \rho \beta} - \, T^{\lambda \alpha}{}_\lambda \, g^{\rho \beta} \right).
\label{galstar3}
\ee
Defining the tensor
\be
S^{\rho\mu\nu} = - S^{\rho\nu\mu} :=  K^{\mu\nu\rho} -g^{\rho\nu}T^{\sigma\mu}{}_\sigma + g^{\rho\mu} T^{\sigma\nu}{}_\sigma  ,
\label{S}
\ee
the generalized Hodge dual torsion assumes the form
\be
\star T^\rho{}_{\mu \nu} = \frac{h}{2} \, \epsilon_{\mu \nu \alpha \beta} \, S^{\rho\alpha\beta}.
\label{galstar4}
\ee
We remark that solutions (\ref{solution1}) and (\ref{solution2}) are the only ones that make the dual torsion to explicitly depend on the contortion tensor.

\subsubsection{Curvature}

Let us consider now the curvature tensor. Analogously to the torsion case, we define its generalized dual by taking into account all possible contractions,
\ba
\star R^{\alpha \beta}{}_{\mu \nu} = h \, \epsilon_{\mu \nu \rho \sigma} \Big[
a \, R^{\alpha \beta \rho \sigma} + b ( R^{\alpha \rho \beta \sigma} -
R^{\beta \rho \alpha \sigma}) \nonumber \\
\qquad \qquad \qquad ~~ + c ( g^{\alpha \rho} \, R^{\beta \sigma} - g^{\beta \rho} \, R^{\alpha \sigma} ) +
d \, g^{\alpha \rho} \, g^{\beta \sigma} \, R \Big],
\label{galstar5}
\ea
with $a, b, c, d$ constant coefficients. We remark that the anti-symmetry in $\alpha$ and $\beta$ is necessary because the curvature 2-form takes values on the Lie algebra of the Lorentz group. By requiring that
\be
{\star \star}R^{\alpha \beta}{}_{\mu \nu} = - R^{\alpha \beta}{}_{\mu \nu},
\ee
we obtain a system of algebraic equations for $a, b, c, d$, whose unique solution is
\be
a = 1/2 \quad \mbox{and} \quad b = c = d = 0.
\ee
This means that for curvature the generalized Hodge dual coincides with the usual definition, that is,
\be
\star R^{\alpha \beta}{}_{\mu \nu} = \frac{h}{2} \, \epsilon_{\mu \nu \rho \sigma} \, 
R^{\alpha \beta \rho \sigma}.
\label{galstar6}
\ee

\section{An application: gravitational lagrangian}

Teleparallel gravity \cite{tg} is characterized by the vanishing of the spin connection:\footnote{All magnitudes related to teleparallel gravity will be denoted with an over ``$\bullet$''.} $\Aw^a{}_{b\mu}=0$. The curvature and torsion tensors in this case are given respectively by
\be
\Rw^a{}_{b \nu \mu} = 0 \quad \mbox{and} \quad
\Tw^a{}_{\nu \mu} = \partial_{\nu} h^{a}{}_{\mu} -
\partial_{\mu} h^{a}{}_{\nu}.
\ee
Through a contraction with a tetrad, the torsion tensor assumes the form
\be
\Tw^\rho{}_{\nu \mu} = \Gammaw^{\rho}{}_{\mu \nu} - \Gammaw^{\rho}{}_{\nu \mu},
\label{TorDef}
\ee
where
\be
\Gammaw^{\rho}{}_{\nu \mu} = h_{a}{}^{\rho} \partial_{\mu} h^{a}{}_{\nu}
\ee
is the Weitzenb\"ock connection. It can be decomposed in the form
\be
\Gammaw^{\rho}{}_{\nu \mu} = \Gammabol^{\rho}{}_{\nu \mu} + \Kw^{\rho}{}_{\nu \mu},
\label{gecow}
\ee
with $\Kw^{\rho}{}_{\nu \mu}$ the contortion of the Weitzenb\"ock torsion.

Now, teleparallel gravity corresponds to a gauge theory for the translation group \cite{gemt}. As such, its action is given by \cite{fadnov}
\begin{equation}
\Sw = \frac{1}{ck} \int \, \eta_{ab} \, \Tw^a \wedge {\star}\Tw^b,
\label{action1}
\end{equation}
where $k = 16 \pi G/c^4$ and
\begin{equation}
\Tw^a = \textstyle{\frac{1}{2}} \, \Tw^a{}_{\mu\nu} \, \rmd x^\mu \wedge \rmd x^\nu
\quad \mbox{and} \quad
{\star}\Tw^a = \textstyle{\frac{1}{2}} \, \star\Tw^a{}_{\rho \sigma} \, \rmd x^\rho \wedge
\rmd x^\sigma 
\label{Tforms}
\end{equation}
are respectively the torsion 2-form and the corresponding dual form. Substituting these expressions in equation~(\ref{action1}), it becomes
\begin{equation}
\Sw =
\frac{1}{4 c k} \int \, \eta_{ab} \, \Tw^a{}_{\mu\nu} \;
\star\Tw^b{}_{\rho \sigma} \, \rmd x^\mu \wedge \rmd x^\nu \wedge \rmd x^\rho \wedge \rmd x^\sigma.
\label{action3}
\end{equation}
Using the identity
\begin{equation}
\rmd x^\mu \wedge \rmd x^\nu \wedge \rmd x^\rho \wedge \rmd x^\sigma = - \,
\epsilon^{\mu \nu \rho \sigma} \, h^2 \, \rmd^4x,
\end{equation}
the action functional reduces to
\begin{equation}
\Sw = -
\frac{1}{4 c k} \int \, \Tw_{\alpha\mu\nu} \;
\star\Tw^{\alpha}{}_{\rho \sigma} \, \epsilon^{\mu \nu \rho \sigma} \; h^2 \, \rmd^4x.
\label{action4}
\end{equation}
Using then the generalized dual definition (\ref{galstar4}), as well as the identity
\be
\epsilon^{\mu \nu \rho \sigma} \epsilon_{\alpha \beta \rho \sigma} =
- \, \frac{2}{h^2} \left(\delta_{\alpha}^{\mu} \delta_{\beta}^{\nu}
- \delta_{\alpha}^{\nu} \delta_{\beta}^{\mu} \right).
\ee
we get
\begin{equation}
\Sw =
\frac{1}{2ck} \int \Tw_{\rho \mu\nu} \,
{\stackrel{\bullet}{S}}{}^{\rho \mu\nu} \, h \, \rmd^4x.
\label{TeleAction}
\end{equation}
This action yields the lagrangian
\be
\Lw =
\frac{h}{2k} \, \Tw_{\rho \mu\nu} \,
{\stackrel{\bullet}{S}}{}^{\rho \mu\nu},
\label{TeleLa}
\ee
which is precisely the lagrangian of the teleparallel equivalent of general relativity~\cite{maluf94}. Using equations (\ref{TorDef}) and (\ref{gecow}), a straightforward calculation shows that it can be rewritten in the form
\be
\Lw = - \frac{h}{k} \, \Rbol - \partial_\mu \left(\frac{2 h}{k} \;
\Tw^{\nu \mu}{}_\nu \right).
\ee
Up to a divergence, therefore, the lagrangian of a gauge theory for the translation group with the Hodge dual given by equation~(\ref{galstar3}) yields the Einstein--Hilbert la\-gran\-gi\-an of general relativity. This shows the consistency --- and actually the necessity --- of the generalized Hodge dual definition (\ref{galstar4}).

\section{Final remarks}

For soldered bundles, the Hodge dual must be generalized in order to take into account all additional contractions allowed by the presence of the solder form. Although for curvature the generalized dual operation turns out to coincide with the usual one, for torsion it gives a completely new dual definition. The importance of this new definition can be verified by analyzing several aspects of gravitation. For example, starting from the standard lagrangian of a gauge theory for the translation group, it naturally yields the lagrangian of the teleparallel equivalent of general relativity, and consequently also the Einstein-Hilbert lagrangian of general relativity. That is to say, it connects the Einstein-Hilbert lagrangian with a typical gauge lagrangian. It is important to remark that the generalized Hodge dual (\ref{galstar4}) has already been used previously \cite{dual1}, but it was guessed just to yield the desired result. Here we have shown that it can be obtained in a constructive way from first principles.

\appendix
\section{Torsion decomposition and the dual}

As is well known, torsion can be decomposed in irreducible components under the global Lorentz group~\cite{hb73}. In terms of these components it reads
\begin{equation}
T_{\lambda \mu \nu} = \textstyle{\frac{2}{3}} \left(t_{\lambda \mu \nu} -
t_{\lambda \nu \mu} \right) + \frac{1}{3} \left(g_{\lambda \mu} v_\nu -
g_{\lambda \nu} v_\mu \right) + \epsilon_{\lambda \mu \nu \rho} \, a^\rho,
\label{deco}
\end{equation}
where
\begin{equation}
v_{\mu} =  T^{\nu}{}_{\nu \mu} \quad \mbox{and} \quad a^{\mu} = \textstyle{\frac{1}{6}}\,  \epsilon^{\mu\nu\rho\sigma} \, T_{\nu\rho\sigma}
\label{pt3}
\end{equation}
are respectively the vector and axial vector parts, and
\begin{equation}
t_{\lambda \mu \nu}  = \textstyle{\frac{1}{2}} \left(T_{\lambda \mu \nu} +
T_{\mu\lambda \nu} \right) + \frac{1}{6} \left(g_{\nu \lambda} v_\mu +
g_{\nu \mu} v_\lambda \right) - \frac{1}{3} g_{\lambda \mu} \, v_\nu,
\label{pt1}
\end{equation}
is a purely tensor part, that is, a tensor with vanishing vector and axial torsions. Using the generalized dual definition (\ref{galstar1}), a simple calculation shows that
\be
\star v_\mu = - 6 h (a - b) \, a_\mu \equiv A \, h \, a_\mu
\label{DualVecTor}
\ee
and
\be
\star a_\mu = \frac{1}{3 h} (2 a + b + 3 c ) \, v_\mu \equiv \frac{B}{h} \, v_\mu,
\label{DualAxiTor}
\ee
where $A$ and $B$ are two new parameters which, on account of the property~(\ref{relation}), satisfy the relation $A \, B = - 1$. In terms of the irreducible components, the generalized dual torsion is easily seen to be
\be
\star T^\lambda{}_{\mu \nu} = h \Big[\pm \frac{2}{3} \, \epsilon_{\mu \nu \alpha \beta} \,
t^{\lambda \alpha \beta} + \frac{A}{3} ( \delta^\lambda{}_{\mu} a_\nu -
\delta^\lambda{}_{\nu} a_\mu) + \frac{B}{h^2} \, \epsilon^\lambda{}_{\mu \nu \rho} \, v^\rho \Big].
\ee
We see from this expression that two parameters suffice to define the generalized dual.

\ack
The authors would like to thank R. Aldrovandi for useful discussions. They would like to thank also FAPESP, CAPES and CNPq for partial financial support.


\end{document}